\documentclass[two column,prl]{revtex4}
\usepackage{bm}
\usepackage{graphics}
\usepackage{graphicx}
\usepackage{amsfonts}
\usepackage{amsmath}
\usepackage{amssymb}
\usepackage{graphicx}

\begin{document}

\preprint{HEP/123-qed}
\title[Short title for running header]{High resolution SIMS depth profiling of nanolayers}
\author{S.V. Baryshev}
\email{sergey.v.baryshev@gmail.com} \affiliation{Material Science
Division, Argonne National Laboratory, 9700 S. Cass Ave., Argonne,
IL 60439}
\author{A.V. Zinovev}
\affiliation{Material Science Division, Argonne National
Laboratory, 9700 S. Cass Ave., Argonne, IL 60439}
\author{C.E. Tripa}
\affiliation{Material Science Division, Argonne National
Laboratory, 9700 S. Cass Ave., Argonne, IL 60439}
\author{M.J. Pellin}
\affiliation{Material Science Division, Argonne National
Laboratory, 9700 S. Cass Ave., Argonne, IL 60439}
\author{Q. Peng}
\affiliation{Energy Systems Division, Argonne National Laboratory,
9700 S. Cass Ave., Argonne, IL 60439}
\author{J.W. Elam}
\affiliation{Energy Systems Division, Argonne National Laboratory,
9700 S. Cass Ave., Argonne, IL 60439}
\author{I.V. Veryovkin}
\email{verigo@anl.gov} \affiliation{Material Science Division,
Argonne National Laboratory, 9700 S. Cass Ave., Argonne, IL 60439}


\begin{abstract}
We report results of high-resolution TOF SIMS (time of flight
secondary ion mass spectrometry) depth profiling experiments on a
nanolayered structure, a stack of 16 alternating MgO and ZnO
$\sim$5.5 nm layers grown on a Si substrate by atomic layer
deposition. The measurements were performed using a newly
developed approach implementing a low energy direct current
normally incident Ar$^+$ ion beam for sample material removal by
sputtering (250 eV and 500 eV energy), in combination with a
pulsed 5 keV Ar$^+$ ion beam at 60$^\circ$ incidence for TOF SIMS
analysis. By this optimized arrangement, a noticeably improved
version of known dual-beam (DB) approach to TOF SIMS depth
profiling is introduced, which can be called \emph{gentle}DB.

We apply the mixing-roughness-information model to detailed
analysis of experimental results. It reveals that the
\emph{gentle}DB approach allows ultimate depth resolution by
confining the ion beam mixing length to about 2 monolayers. This
corresponds to the escape depth of secondary ions, the fundamental
depth resolution limitation in SIMS. Other parameters deduced from
the measured depth profiles indicate that a single layer thickness
equals to 6 nm so that "flat" layer thickness $d$ is of 3 nm and
interfacial roughness $\sigma$ is of 1.5 nm thus yielding
$d$+2$\cdot\sigma$=6 nm. In essence, we have demonstrated that the
\emph{gentle}DB TOF SIMS depth profiling with noble gas ion beams
is capable of revealing structural features of a stack of
nanolayers, resolving its original surface and estimating the
roughness of interlayer interfaces, which is difficult to obtain
by traditional approaches.

\end{abstract}

\maketitle

\section{Introduction} Secondary ion mass spectrometry, SIMS, is a well-established
analytical method in surface, materials and chemical sciences.
Besides conducting routine elemental, molecular, and isotopic
analyses of solid surfaces and thin films, creative applications
of this method help in finding solutions to a variety of
non-standard materials science problems \cite{1,2,3,4,5}. Sputter
depth profiling analysis provides information on the variation of
sample composition with depth below the initial surface. In such
applications, the SIMS method is in a class of its own due to the
unique combination of high sensitivity and depth resolution. While
for other surface sensitive analytical techniques, such as Auger
Electron or X-ray Photoelectron Spectroscopies (AES or XPS), the
inherent escape depth is $\gtrsim$1 nm, for SIMS this
characteristic parameter on the order of 1 to 2 atomic layers,
which can be considered as a theoretical limit of its depth
resolution \cite{6,7}. In real life, however, the depth resolution
of SIMS is limited by an native surface roughness of the sample
(if any) and by the fundamental effect called ion beam mixing
\cite{8}. Mixing is characterized by the ion beam mixing length
$w$, which is a figure-of-merit of a SIMS instrument. For all
depth profiling analyses, the former limiting factor is out of the
analyst's control. Fortunately, one can minimize the latter effect
by optimizing analytical approaches. Often, new combinations of
techniques for sample material erosion and composition probing are
sought after. As an alternative to sputtering, sample material can
be removed for compositional probing by laser enhanced field
desorption \cite{9,10}, as implemented in modern atom probe
tomography instruments. In the case of SIMS depth profiling
analysis, all such approaches, in general, aim at decreasing the
energy deposited into the sample by incident projectiles. This can
be achieved by either directly decreasing the ion beam energy, as
usually done for conventional primary ions (Cs$^+$, noble gas
ions, and O$_2^+$ as well) \cite{11,12}, or implicitly, by having
more constituent atoms in the primary molecular/cluster ions
(single charged SF$_6$ \cite{13}, C$_{60}$ \cite{14} molecules/Au
and Bi \cite{15}, or Ar \cite{16} clusters) and dividing this
energy between the constituents so that each atom has much lower
impact energy ($\sim$1/60 of the nominal beam energy per single
constituent atom in the case of C$_{60}$).

\par The dual-beam (DB) depth profiling \cite{17} was introduced in time of
flight (TOF) SIMS instruments to decouple the depth resolution of
the direct current (dc) sputtering ion beam from the lateral
resolution of the pulsed analysis ion beam. In the DB approach,
the sputtering and analysis cycles are independent, and the depth
resolution is controlled only by the sputtering beam parameters
\cite{18}. In order to fully realize the potential of the DB
sputter depth profiling approach, we further improved it by
implementing a normally incident sputtering ion beam, which is
extracted from the ion source at nominal energy of a few keV and
delivered with the same energy into the target region, where the
target potential can decelerate it so that the impact energy does
not exceed a few hundred eV \cite{19}. This permits direct
adjustment of the sputtering beam energy (down to the sputtering
threshold) without losses of the ion beam current and degradation
of the beam spot quality. An important benefit of the normal
incidence sputtering is the significant reduction in roughening of
the surface induced by the ion beam \cite{20}. Moreover, the lower
the impact energy, the lower the ion mixing effects are. Thus, the
low energy normally incident sputtering ion beam maximizes the
benefits of the DB sputter depth profiling and allows us to create
a versatile instrument, suitable both for ultra-trace surface
analysis \cite{19,21} and applications in the materials science
applications, and thus address a current problem on successful
mass application of nanodevices discussed in Ref.\cite{22}. To
resolve this problem, development of advanced and powerful
characterization tools, combining multiple points of view (i.e.
probes) on a sample, is highly needed to achieve \emph{faster,
facile and standardized characterization} and so to \emph{reap the
benefits of nanotechnology} \cite{22}.

\par We apply here \emph{gentle}DB approach to better understand
chemistry and structure of ultrathin films produced by atomic
layer deposition (ALD). The ALD technique is widely used for a
layer-by-layer materials synthesis and has great application
potential in many areas due to its ability to coat
high-aspect-ratio substrates by conformal layered structures with
precisely predefined compositional profiles \cite{23,24}. One of
such important application areas is functionalization of material
surfaces for use in novel detectors and sensors. Our effort here
is the contribution to the Large Area Picosecond Photo-Detector
(http://psec.uchicago.edu/) collaboration, which is focused on
fast particle detectors with large areas functionalized by ALD. It
is known that thin films of metal oxides can serve as very
efficient emitters of secondary electrons and be used for
improving detector performance. Moreover, it has been recently
shown that such ALD-grown films of several nm thickness enable
independent tailoring of the electrical resistance and the
secondary electron emission yield \cite{25}. A functionalization
of a detector based on a large area (tens square centimeters)
microchannel plate means that surfaces of the vast amount of
microchannel plate pores must be coated by thin films with
predefined characteristics. This task is the perfect match to the
unique capabilities of ALD. At the same time, understanding and
controlling the coating uniformity (thickness fluctuation etc.)
and the condition of interfaces between resistive and emitting
layers as well as between those and the substrate are extremely
important for improving the materials' performance and
functionality.

\par With this goal in mind, we characterize in this work a stack of
16 alternating $\sim$5.5 nm MgO and ZnO layers (8 of each) using
\emph{gentle}DB SIMS depth profiling approach with normally
incident 250 eV and 500 eV Ar$^+$ ion beams for sputtering and a
pulsed 5 keV Ar$^+$ beam (pointed at 60$^\circ$ from the target
normal) for analysis. The chosen stack of 8 pairs MgO/ZnO, being
one of the proposed functionalization systems, serves two
purposes. The first one is to create a planar (unfolded) model of
an actual microchannel plate pore surface coating with complex
geometry coating in order to reveal the layer-to-layer
reproducibility of the ALD process starting from the substrate,
along with the layers uniformity and the conditions of interlayer
interfaces. The other purpose is to test the depth profiling
capabilities of the \emph{gentle}DB approach. Besides, multilayer
nanostructures and devices are at the forefront of materials
science. Giant magnetoresistance multilayer films \cite{26} or
superlattices of transition metal oxides \cite{27} are striking
examples of this. Desired performance of such structures strongly
depends on the interfacial roughness, interdiffusion between
layers, layer-to-layer consistency, and layer conformality as
well. Thus, we aim to demonstrate in this work the unique synergy
of the combination of the low energy sputtering with inert gas
ions and the normal incidence angle, which makes the
\emph{gentle}DB depth profiling approach superior to many others,
especially when applied to characterization of nanolayers. In this
case, the capability of sputter depth profiling without alteration
of the sample chemical composition and modification of its surface
and interface morphology does minimize instrumental artifacts and
experimental/procedural assumptions and thus helps to elucidate
the relationship between these parameters and advance materials
synthesis approaches. We will support these statements by
demonstrating accurate depth profiling on the nanometer scale and
determining key characterization parameters such as the ion beam
mixing length, and the decay $\lambda_d$ and leading
$\lambda_{up}$ lengths, which effectively characterize the
resolution of an instrument and can be compared with those of
other approaches/setups in the field. The proposed
characterization approach becomes even more important if one
imagines that less-than-nm depth resolution is achieved laterally
over the area of $\sim$1 mm$^2$ (in general, this area can be
varied between $\sim$100$\times$100 $\mu$m$^2$ to several mm$^2$).

\section{Samples} A layered structure $|$5.5 nm MgO/5.5 nm ZnO$|$$\times$8 was grown by the ALD
technique on a Si substrate and characterized by x-ray diffraction
and ellipsometry. The standard calibration and characterization
procedures of ALD can be found in Refs.\cite{24,28,29,30}. The
structural data obtained from these measurements indicated that
MgO layers are amorphous and ZnO layers are polycrystalline in the
wurtzite phase. Although the surface roughness of the MgO/ZnO
sample was not measured, Al$_2$O$_3$/ZnO multilayers prepared
under similar conditions had a surface roughness of 0.9-1.5 nm
\cite{24,30}. The roughness of the Si substrate was $\sim$0.3 nm.
The initial layer-to-layer mixing due to thermal diffusion during
growth at $T$=473 K is expected to be extremely low.

\section{Experimental} The depth profiling of the layered structure
described above was performed using a custom made SARISA (Surface
Analysis by Resonance Ionization of Sputtered Atoms) instrument
operated in SIMS mode (instead of its primary mode of secondary
neutral mass spectrometry with laser post-ionization)
\cite{19,31}. The SARISA TOF mass spectrometer was designed to
operate in multiple modes using the same set of ion optics
\cite{19}. With post-ionization lasers switched off, the
instrument operation corresponds to that of TOF SIMS with long
primary ion pulses ($\sim$0.2-1 $\mu$s) and delayed extraction of
secondary ions. For the experiments described below, SARISA's ion
optics either delivered the primary sputtering Ar$^+$ ion beam to
the sample assuring its orthogonal impact onto sample surface and
controlling its impact energy, or it extracted secondary ions
generated by a pulsed analytical Ar$^+$ ion beam and performed
their TOF MS analysis \cite{19}. Fast switching between these
regimes was done electronically by changing the potentials of the
optics electrodes.

\par The low energy ion beam is formed by injecting a keV ion beam
produced by a VG EX05 gun into the TOF spectrometer ion optics, by
delivering and focusing the beam at normal incidence to the sample
and by decelerating the ions to the desired low impact energy by
biasing the target with an appropriate voltage (see Fig.1). The
beam defocusing, which accompanies this deceleration, is
compensated by changing the voltage of the outlet electrostatic
lens called Lens1 \cite{19}. The impact energy of this milling
beam was set in this study to either 250 or 500 eV, and the ion
current was $\sim$1 $\mu$A. This beam was raster scanned over the
sample surface by engaging an octupole deflector present in the
TOF system ion path (the Shaping Octupole in Fig. 1), so that a
square crater of $\sim$1.5$\times$1.5 mm$^2$ (shown in Fig.1 as a
blue square) was eroded. The exact size of the crater depends on
the deceleration potential of the target, which additionally
deflects the beam when it is raster scanned. For example, the
ratio of the crater sides between $\varepsilon_b$=250 eV and
$\varepsilon_b$=3 keV equals 1.3; this value was proved both by
crater imaging and by SIMION 3D$^\copyright$ \cite{32} simulation.

\par The analytical ion beam, used to probe the sample at various
depths, comes from an Atomika WF421 gun pointed at 60$^\circ$
incidence with respect to the target normal (Fig.1). This ion beam
can be independently tuned, raster scanned, and pulsed. The
primary Ar$^+$ ion energy was always 5 keV, and the pulse duration
used in the analysis in all the experiments was 200 ns, since mass
resolution under these conditions was sufficient. The raster scan
size of this ion beam in all the experiments described here was
set to a 500$\times$500 $\mu$m$^2$ (the green square in Fig.1).
During the TOF MS analysis cycle, the deceleration potential on
the target is switched off. In the \emph{gentle}DB depth profiling
regime, the Atomika ion beam was only used for analysis. By
choosing an appropriate aperture, the primary ion current was set
to 30 nA, in order to attenuate the secondary ion signal so that
the TOF MS detector is not overwhelmed.

\par For the two ion beams used in these experiments, parameter
$\alpha=E_{Atomika}/E_{low-energy}$ was 10$^{-5}$ \cite{18}.

\par In order to perform the \emph{gentle}DB depth profiling, we fit
the analysis beam raster into the center of the low energy
sputtering beam raster. To do this pre-alignment, we used a
Faraday cup for ion beam collection and took advantage of
available crater imaging capabilities. Crater imaging was done
\emph{in situ} by an optical Schwarzschild-type microscope
\cite{19} and the SEM \cite{19,33} or by \emph{ex situ} optical
white light interferometry \cite{34} (Fig.1b).

\begin{figure}[t] \centering
\includegraphics[width=8cm]{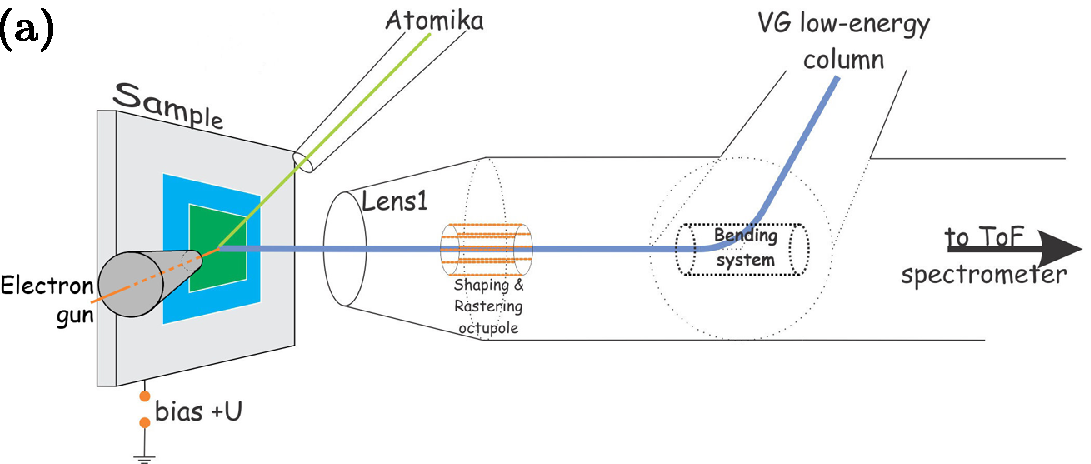}

\

\includegraphics[width=6cm]{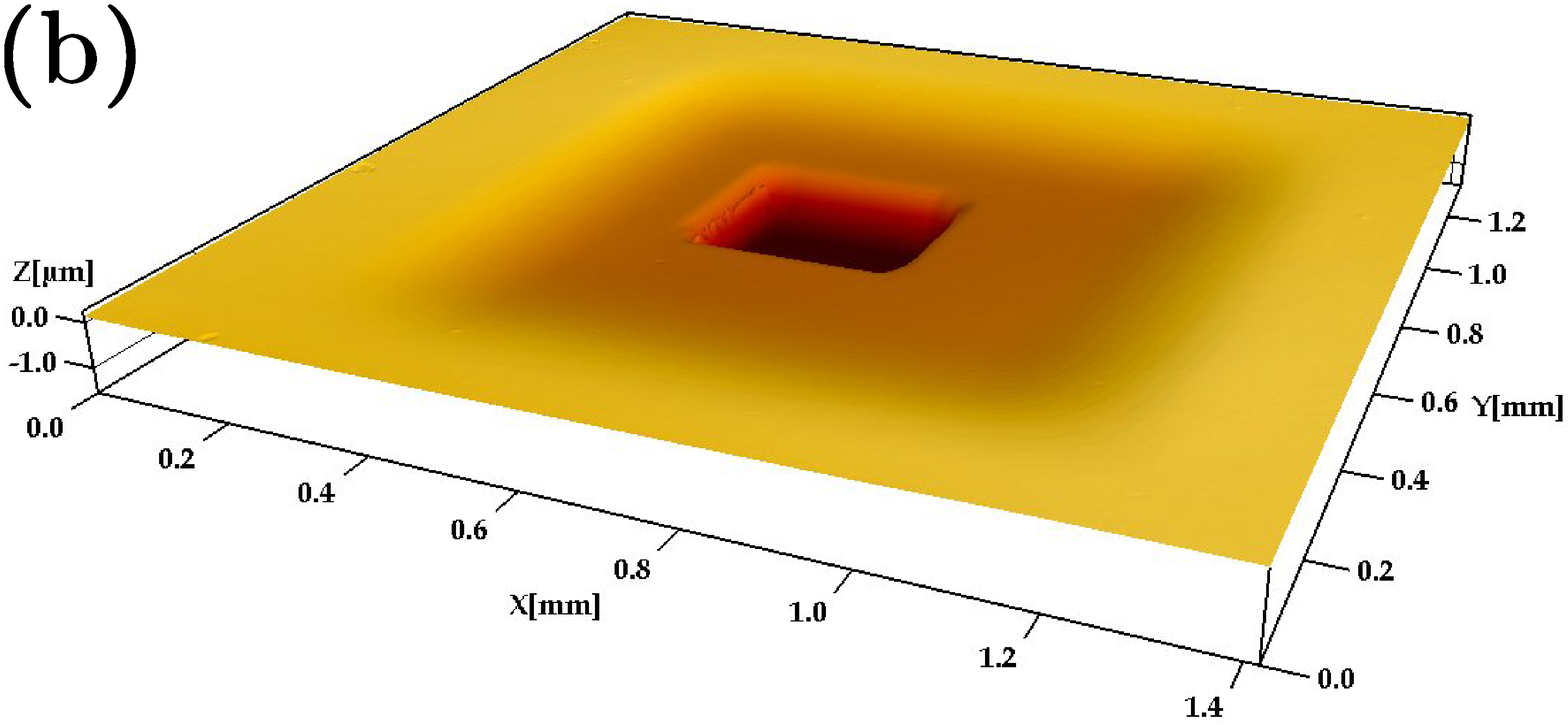}
\caption{(a) Diagram of the dual-beam system. The primary Ar$^+$
beam comes from the VG low energy column and is deflected into the
front TOF column by the Bending System optics \cite{19}. The ion
beam is tightly focused with the front end Lens1, impinging at
normal incidence with respect to the sample surface. The beam is
raster scanned by the Shaping Octupole. The separate Atomika ion
gun is pointed at the target at 60$^\circ$ from its normal and
equipped with a pulsing and raster scanning plates. The electron
gun images surface topography. (b) White light interferometry
image of a high-energy Atomika Raster crater positioned in the
center of a low-energy VG Raster crater. The Atomika Raster crater
was dc sputtered just after the completion of the \emph{gentle}DB
SIMS measurements in order to reveal the place where the analysis
was done, since the pulsed analytical beam does not produce any
visible bowl under conditions satisfying $\alpha\ll 1$.}
\end{figure}

\section{Results and Discussion} Figure 2 demonstrates normalized
depth profiles obtained by the \emph{gentle}DB approach at 250 and
500 eV sputtering beam energies. Both the $^{24}$Mg and $^{64}$Zn
depth profiles are very similar and feature high contrast and high
stability of the signal all the way through the multilayer
structure. The model structure of the $^{24}$Mg distribution based
on ellipsometry measurements is shown in Fig.2 as a black dotted
line. The end of the last layer depth profile is marked by a
$^{28}$Si signal spike from the substrate. The variation in the
$^{28}$Si SIMS signal is due to variation in the Si secondary ion
yield $\gamma$ from SiO$_2$, and its following drop and
stabilization are observed at the 100-fold lower level with
respect to the spike magnitude, which coincides with the known
ratio $\frac{\gamma_{SiO_2}}{\gamma_{Si}}\sim 100$. The first peak
in the $^{24}$Mg profile has an unsymmetrical shape
(cotangent-like) starting at the surface from the highest SIMS
intensities transforming right into trailing edge, while the
$^{64}$Zn profile starts approximately at the zero level; both
profiles then have a steady sine-like shape. Table I shows the
leading $\lambda_{up}$ and decay $\lambda_{d}$ lengths extracted
from the depth profiles. The definition for the leading or decay
length is the inverse of the first order coefficient for the
linear approximation of the leading or trailing edges of a peak
represented in semi-log scale, $\ell n$(Intensity) vs. depth
\cite{35}. Here, both $\lambda_{up}$ and $\lambda_{d}$ are $\sim$5
nm for $^{24}$Mg, $\sim$2 nm for $^{64}$Zn, and $\lambda_{up}$=0.4
nm for $^{28}$Si. The measured leading and decay lengths were on
the same order or better than published to-date corresponding data
obtained with commercial state-of-the-art SIMS instrumentation
using cluster/molecular or low energy O$_2^+$/Cs$^+$ ion sources,
or with laser-assisted atom probe tomography
\cite{9,10,12,13,17,36,37}.

\begin{figure}[t] \centering
\includegraphics[width=7cm]{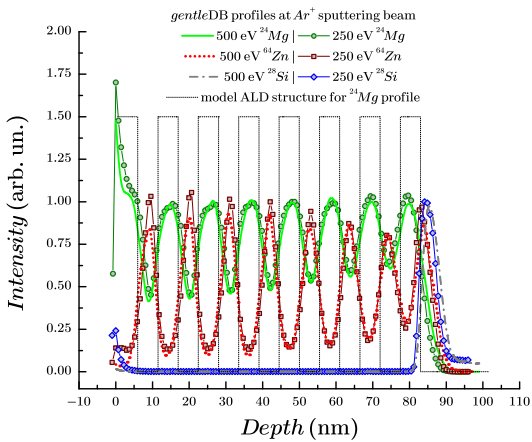}
\caption{SIMS depth profiles of the MgO/ZnO layers on a Si
substrate in the \emph{gentle}DB approach obtained by a 250 eV and
500 eV sputtering Ar$^+$ ion beam combined with a 5 keV analysis
Ar$^+$ ion beam. The profile of the first MgO layer is poorly
resolved because of surface contamination with Mg.}
\end{figure}

\newpage

\begin{center}
Table I. Leading and decay lengths for $^{24}$Mg, $^{64}$Zn, and
$^{28}$Si extracted from \emph{gentle}DB depth profiles.
\begin{tabular}{c|c c c|c c c}
  \hline
   &  & 500 eV &  &  & 250 eV &  \\
  \hline
  (nm)                          & $^{24}$Mg & $^{64}$Zn & $^{28}$Si & $^{24}$Mg & $^{64}$Zn & $^{28}$Si \\
  \hline
  Average $\lambda_{up}$        & 5.6       & 2.1       & --        & 4.9       & 2.2       & --        \\
  \hline
  Average $\lambda_{d}$         & 5.1       & 2.1       & --        & 4.7       & 2.0       & --        \\
  \hline
  $\lambda_{d}$ at Si interface & 1.3       & 1.1       & --        & 0.9       & 0.8       & --        \\
  \hline
  Si $\lambda_{up}$             & --        & --        & 0.4       & --        & --        & 0.4       \\
  \hline
\end{tabular}
\end{center}

\par As mentioned above, the first $^{24}$Mg peak of the stack was
not resolved from the surface constituent, which caps the
structure. In order to resolve in the depth profile the $^{24}$Mg
peak corresponding to the topmost ALD layer from the surface
contamination, we decreased the time of the individual sputtering
cycles and proportionally increased their number. Figure 3 shows
the depth profile of the two top layers performed with more data
points per depth unit to separate the 24 mass from the topmost
layer $^{24}$Mg peak. The green line here corresponds to the depth
profile with poorer resolved Mg profile, shown in Fig.2. This is
to emphasize the capacity in the resolving power tuning of the
\emph{gentle}DB approach. The distance between subsequent depth
points in Fig.3 is $\sim$0.2 nm, which precisely correlates with
half of the average lattice parameter of 3.9 ${\AA}$ for wurtzite
ZnO \cite{38} and half of the cubical unit cell size of
crystalline MgO with a lattice parameter of 4.2 ${\AA}$ \cite{39}.

\par The ratios of $^{25}$Mg/$^{24}$Mg and $^{26}$Mg/$^{24}$Mg
proved that the mass 24 peak appears to be Mg with correct
terrestrial isotopic ratio values. Such stoichiometry enrichment
may be due to either natural surface contamination occurring after
the sample is taken out of the ALD reactor, or due to some ALD
specific effect manifesting itself at finalizing growth steps. At
the same time, the high depth resolution profiling allows us to
detect a "dip" between the surface $^{24}$Mg and the topmost
magnesia layer so that there is enough prior information to
precisely reconstruct the contaminant and MgO layer profiles,
assuming the contamination is natural. These two functions are
displayed in Fig.3: cyan dotted line corresponds to MgO, while
black dotted one is the surface Mg cap.

\begin{figure}[t] \centering
\includegraphics[width=7cm]{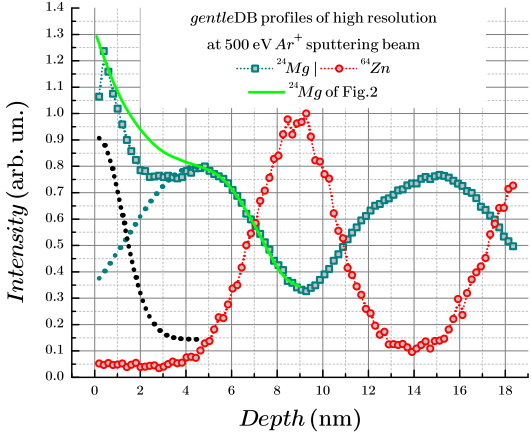}
\caption{High-resolution depth profiles of MgO/ZnO top layers at
500 eV sputtering Ar$^+$ beam and 5 keV analytical Ar$^+$ beam
showing the surface Mg contamination being resolved from the top
MgO layer peak. The green solid line is the starting part of the
profiles copied from Fig.2. Cyan and black dotted lines are the Mg
signals reconstructed from $^{24}$Mg high resolution depth profile
(open squares + dotted line in cyan) curve which correspond to
$^{24}$Mg in ALD magnesia and some sort of contaminant,
respectively.}
\end{figure}

\par These experimental results clearly demonstrate remarkable
resolving power of the \emph{gentle}DB approach to depth
profiling. In this regard, keeping in mind that normal incidence
should not affect pronouncedly native sample roughness \cite{20},
let us qualify what the definition "gentle" means. For this, we
apply the MRI (mixing-roughness-information) model \cite{40,41} to
simulate the $^{64}$Zn peak in the 500 eV high resolution
\emph{gentle}DB depth profiles using the model parameters $w$
(cascade mixing length), $\sigma$ (r.m.s., root mean square,
roughness), and $\lambda$ (information depth, or escape depth,
$\sim$1--2 monolayer in the case of SIMS).

\par A summary of the MRI model for the SIMS experiment is shown
below. Note that the preferential sputtering effect is not taken
into account here, since there is no such an effect for MgO and
ZnO. As a starting point, the ion mixing is modeled first by the
differential equation (1) for an ideal, flat layer of thickness
$d$, with the boundary conditions (2), (4), and (5)

\[
\frac{dC_{M}(z)}{dz}+\frac{1}{w}\cdot C_{M}(z)=\frac{1}{w}\cdot
C_{act}(z+w) \text{. (1)}
\]

\[
\left\{
\begin{array}{l l}
  C_{M}^{up}(z)=A\cdot(1-exp(-\frac{z}{w}+\alpha) \\
  C_{M}^{up}(z)|_{z_1-w}=0 \\
\end{array} \text{. (2)}\right.
\]
Hence,
\[
\left\{
\begin{array}{l l}
  \alpha=\frac{z_1-w}{w} \\
  A=1 \text{ by default} & \text{. (3)} \\
  C_{M}^{up}(z)=1-exp(-\frac{z-z_1+w}{w}) \\
\end{array} \right.
\]

\[
C_{act}(z+w)|_{z\geq z_2-w}=0 \text{. (4)}
\]

\[
\left\{
\begin{array}{l l}
  C_{M}^{d}(z)=B\cdot exp(-\frac{z}{w}+\beta) \\
  C_{M}^{d}(z)|_{z_2-w}=C_{M}^{up}(z)|_{z_2-w} \\
\end{array} \text{. (5)}\right.
\]
Hence,
\[
\left\{
\begin{array}{l l}
  \beta=\frac{z_2-w}{w} \\
  B=1-exp(-\frac{z_2-z_1}{w}) & \text{, (6)} \\
  C_{M}^{d}(z)=(1-exp(-\frac{z_2-z_1}{w}))\cdot exp(-\frac{z-z_2+w}{w}) \\
\end{array} \right.
\]
where $C_M(z)$ and $C_{act}(z+w)$ are apparent (at any depth $z$)
and actual (at $w$ deeper relative to $z$) concentrations of an
element, respectively; $w$ is the ion mixing length; $z_1$ and
$z_2$ are the flat boundaries of the arbitrary layer of interest,
so that $z_2-z_1=d$ equals the layer thickness, $d$.
$C_{M}^{up}(z)$ and $C_{M}^{d}(z)$ are the leading and trailing
edges of the depth profile peak, respectively. Thus, due to the
cascade mixing, the solution yields
\begin{align*}
&C_M(z)=&                        \\
=       &\left\{
         \begin{array}{l l}
         1-exp(-\frac{z-z_1+w}{w}), z\in[z_1-w; z_2-w]\\
         (1-exp(-\frac{z_2-z_1}{w}))\cdot exp(-\frac{z-z_2+w}{w}), z\geq z_2-w\\
         \end{array} \text{. (7)}\right.
\end{align*}
If the surface roughness (either inherent, or induced, or both) is
taken into account, then the profile additionally broadens as
\[
C_{MR}(z)=\frac{1}{\sigma\cdot\sqrt{2\cdot\pi}}\int_{-\infty}^{\infty}C_M(t)\cdot
exp(-\frac{(z-t)^2}{2\cdot\sigma^2})dt \text{, (8)}
\]
where the mixing is convolved with the Gaussian function that
takes the r.m.s. roughness $\sigma$ as the standard deviation.

\par Finally, when both the $C_{MR}(z)$ parameter and the effective
depth contributing to the signal are considered, the depth profile
shape can then be expressed as
\[
C_{MRI}(z)=\frac{1}{\lambda}\int_{z}^{\infty}C_{MR}(x)\cdot
exp(-\frac{z-x}{\lambda})dx \text{, (9)}
\]
where $\lambda$ is the information depth or ion escape depth, by
conventional definition. Generally speaking, in the case of SIMS,
it has negligible contribution in the elemental peak dependence on
the depth $C_{MR}(z)$, defined only by the cascade mixing and
roughness.

\par We found that the best matching between the measured and
modeled profiles occurs when the MRI parameters $w$ and $\sigma$
were set as follows: $w$=0.4 nm, $\sigma$=1.5 nm, and the
parameter $\lambda$ was fixed at value 0.2 nm corresponding to 1
monolayer. The nominal thickness (flat thickness) of the layer,
$d$, had to be of 3 nm to get the best peak fit in Fig.4. The
thickness $d$ is not a free parameter of the MRI model.
Nevertheless, being unknown a priori, it had to be varied in order
to define the boundary conditions (2), (4), and (5). Then the
effective thickness of a single ZnO layer, i.e., 2$\cdot\sigma+d$,
equals to 6 nm. This value obtained by MRI does agree well with
the ones estimated for a single ALD layer by ellipsometry and
quartz crystal microbalance technique except the real material
layer is not ideally flat, it can be represented by a structure
made of flat inner layer squeezed between comparable in thickness
additions on both sides representing wavy interfaces. Thus, under
\emph{gentle}DB conditions, the value $\sigma$=1.5 nm calculated
for the first $^{64}$Zn peak can be accepted as the native
roughness of the sample inner interfaces (see Fig.4). In agreement
with the MRI model results, this roughness causes symmetrical peak
broadening in sputter depth profiling experiments.

\begin{figure}[t] \centering
\includegraphics[width=7cm]{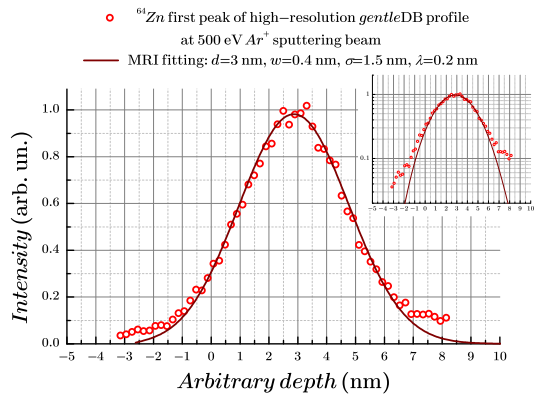}
\caption{MRI model fittings of the first $^{64}$Zn depth profile
peak: 500 eV, high resolution \emph{gentle}DB profile -- red
circle data points vs. the MRI simulated curve in dark red. The
inset is the same plot as the main figure, but in semi-log scale.}
\end{figure}

\par It is interesting, that some preceding works, e.g. Ref.\cite{42},
concluded that the attenuation of the signal intensity of each
subsequent layer in laminate/multilayer structures, by
approximately $\sin(x)\cdot exp(-x)$ law, occurs only due to the
native interface roughness of layered structures. However,
presented here results of the \emph{gentle}DB depth profiling
obtained on a sample of exceedingly disturbed grown interfaces
suggest that the native roughness is not the dominant factor in
the degradation and smearing of depth profiles in SIMS. To prove
this, we applied the conventional single beam TOF SIMS depth
profiling approach to the same sample, using 5 keV Ar$^+$ ions
with 60$^\circ$ incidence angle (generated by the Atomika WF421
ion gun) -- for both ion milling and TOF MS analysis (results to
be published). The comparison between single beam and
\emph{gentle}DB approaches revealed dramatic differences in
results: the ion beam mixing length $w$ increased from 0.4 nm to 3
to 4 nm \cite{43}, which is on the order of the entire ALD layer
thickness. Moreover, we observed with single beam approach the
periodical sine signal with exponential attenuation mentioned
above as well as the smearing of the peak shapes for deeper
layers. To summarize, the ion beam mixing appears to be the most
important phenomenon to account for in depth profiling experiments
on nanolayered structures. Reducing ion mixing length $w$ by a
proper choice of the analytical procedure can produce depth
profiles such as the ones shown in Fig.2 that reveal much more
information about the samples than about the measurement
artifacts/distortions.

\section{Conclusion} To conclude, we have outlined here our approach
to solve the nanomaterial characterization bottleneck problem
\cite{22} in the materials science by application of TOF SIMS
technique. Namely, we described our gentle variant of the
dual-beam sputter depth profiling (\emph{gentle}DB) applied to
characterization of nanolayered materials and demonstrated
sub-nanometer depth resolution of the method. This approach
combines normally incident low energy (down to sputtering
threshold) Ar$^+$ ion beam for sputtering with several keV pulsed
Ar$^+$ ion beam (60$^\circ$ incidence angle) for analysis. The
shallow penetration depth of these sputtering primary ions results
in an ultra-short range of a crystal lattice disturbance of about
0.4 nm thus enabling measurements of structural parameters (flat
thickness and roughness) of a stack of 16 alternating MgO and ZnO
nanolayers (8 of each, with thickness of $\sim$5.5 nm) grown by
ALD on a Si substrate. The interfacial roughness as small as 1.5
nm estimated by the \emph{gentle}DB sputter depth profiling does
agree well with measurements obtained by other techniques: the
surface roughness of individual ZnO layers ($\sim$0.5 nm)
determined by atomic force microscopy \cite{30} and the roughness
of the multilayer structure ($\sim$0.9-1.5 nm) as determined by
x-ray reflectivity and atomic force microscopy for comparable
ALD-grown Al$_2$O$_3$/ZnO multilayers \cite{24,30}.

\textbf{Acknowledgements.} This work was supported under Contract
No. DE-AC02-06CH11357 between UChicago Argonne, LLC and the U.S.
Department of Energy and by NASA through grants NNH08AH761 and
NNH08ZDA001N.

\end{document}